\documentclass[seceq]{ptptex}

\usepackage{epsfig,multicol}
\usepackage{amsmath}

\newcommand{\pslash}{p\kern-1ex /}
\newcommand{\qslash}{q\kern-1ex /}
\newcommand{\lslash}{l\kern-1ex /}
\newcommand{\sslash}{s\kern-1ex /}
\newcommand{\kaslash}{k_a\kern-2ex /}
\newcommand{\kbslash}{k_b\kern-2ex /}
\newcommand{\Dslash}{\mathcal{D}\kern-1.5ex /}

\newcommand{\beqa}{\begin{eqnarray}}
\newcommand{\eeqa}{\end{eqnarray}}

\newcommand{\ba}{\begin{eqnarray}}
\newcommand{\ea}{\end{eqnarray}}
\newcommand{\be}{\begin{equation}}   
\newcommand{\Nf}{N_\mathrm{f}}
\newcommand{\Nfv}{N_{\mathrm{f}}^{\mathrm{v}}}

\title{Toward an understanding of short distance repulsions among 
baryons in QCD\\
-- NBS wave functions and operator product expansion --}
\author{Sinya \textsc{Aoki}$^{1,2}$,  Janos \textsc{Balog}$^3$, Peter \textsc{Weisz}$^4$}

\inst{
$^1$ Graduate School of Pure and Applied Sciences, 
University of Tsukuba, Tsukuba, Ibaraki 305-8571,    Japan\\
$^2$ Center for Computational Sciences, University of Tsukuba, Tsukuba, Ibaraki 305-8577,    Japan\\
$^3$ Institute for Particle and Nuclear Physics, 
Wigner Research Centre for Physics,
H-1525 Budapest 114, P.O.B. 49, Hungary\\
$^4$
Max-Planck-Institut f\"ur Physik, F\"ohringer Ring 6, 
D-80805 M\"unchen, Germany\\
}

\abst{
We report on our recent attempts to determine the short distance 
behaviors of general 2-baryon and 3-baryon forces, 
which are defined from the Nambu-Bethe-Salpeter(NBS) wave function,
by using the operator product expansion 
and a renormalization group analysis in QCD.
We have found that the repulsion at short distance increases 
as the number of valence quarks increases or when the number of 
different flavors involved decreases. This global tendency suggests a 
Pauli suppression principle among quark fields at work.  
}

\begin{document}

\maketitle

\section{Motivation}
The most fundamental quantity in nuclear physics is a force between nucleons, 
the nuclear force, from which major properties of nuclei are extracted. 
The phenomenological nuclear potentials~\cite{NN-review}, 
which well describe nucleon-nucleon (NN) scattering at low energy, 
exhibit long-to-medium distance attractions, 
essential for the binding of atomic nuclei,
as well as short distance repulsion (a repulsive core), 
important for the stability of nuclei against collapse. 
While the former properties have been explained by meson exchanges 
between nucleons~\cite{yukawa}, 
the origin of the repulsive core~\cite{jastrow} 
has not yet been well understood. 
Considerations of the fundamental degrees of freedom for nucleons, 
quarks and gluons, and their dynamics, QCD, 
are required to explain the properties between nucleons 
at such short distances. 

The recent observation of a neutron star as heavy as twice 
a solar mass~\cite{Demorest:2010bx} forces us to consider more general 
short distance repulsions among baryons, not only NN but also 3N, BB and 3B, 
where B represents an octet baryon which may include one or more strange quarks.
This is because although the repulsive core of NN potentials 
is an important ingredient for the determination 
of the maximum mass of neutron stars, 
it alone seems insufficient to sustain the two-solar-mass neutron star.
Moreover, 
an appearance of heavier strange quarks, 
converted from lighter up and down quarks via the weak interaction 
in such a high density environment, makes the equation of state 
softer in the core of the neutron star, 
so that the maximum mass of the neutron star is further reduced. 
It has already been pointed out that general BB potentials alone 
can not sustain the two-solar-mass neutron star~\cite{Nishizaki:2002ih}.
It is also argued that the maximum mass of neutron stars can be increased if 
3-body forces, becoming important at high density, 
also have strong short distance repulsions to compensate 
the reduction due to the appearance of strange quarks.
Furthermore it is thought that such repulsions should be universal: 
they should appear not only in 3NF (3 nucleon forces) 
but also in 3BF (3 baryon forces)~\cite{Nishizaki:2002ih}.
A problem of such an explanation, however, is that
both experimental and theoretical determinations of 
3-body forces have been known to be very difficult at short distances.

Recently, a new method has been proposed 
to define and investigate the NN potential 
from the Nambu-Bethe-Salpeter(NBS) wave function in lattice QCD.
This method was employed successfully to calculate 
NN potentials~\cite{Ishii:2006ec,Aoki:2008hh,Aoki:2009ji},
and it has been widely applied to various hadronic 
interactions~\cite{Ishii:2009zr,Murano:2011nz,Nemura:2008sp,Nemura:2012fm,Sasaki:2010bi,Ikeda:2011qm} 
including BB potentials at $\Nfv=3$~\cite{Inoue:2010hs,Inoue:2010es}, 
and 3N forces at $\Nfv=2$~\cite{Doi:2011gq}, where $\Nfv$ denotes the
number of different quark flavors of the valence quarks 
of the baryonic systems under consideration. 
  
In this paper, we report our recent attempts to determine the short distance 
behaviors of general 2-baryon and 3-baryon forces, 
using the same definition of potentials as employed in 
the lattice QCD simulations referred to above.
In our study we combine the operator product expansion with 
a renormalization group (RG) analysis of QCD using perturbatively 
computed RG functions.
Our results may give not only useful boundary conditions 
of 2- and 3-body forces at short distance 
but also a hint toward an understanding of 
origins for short distance repulsions.
We have observed a tendency toward more short distance repulsions 
for more nucleons: 
In the case of the NN potential, 
the short distance repulsion appears only if the ratio 
of two matrix elements has a positive sign~\cite{Aoki:2010kx,Aoki:2009pi}. 
Although this condition is satisfied at low energy~\cite{Aoki:2010kx}, 
the short distance repulsion in not universal. On the other hand, 
a universal short distance repulsion is shown to appear in 
3N forces~\cite{Aoki:2011aa}.
We also found, for a fixed number of valence quarks,
a weaker short distance repulsion for increasing $\Nfv$.
While the NN potentials (and potentials of general BB systems 
with $\Nfv=2$) have the short distance repulsion at low energy, 
there appear short distance attractions in some channels of BB interactions 
with $\Nfv=3$~\cite{Aoki:2010uz}. 
This tendency is found to be true also in 3B forces. 

In the following, we explain how we obtain the above results and 
what the above statements mean more precisely.
In sect.~\ref{sec:basics}, we first give our definition of potentials in QCD, 
which has been employed successfully in lattice QCD calculations. 
We then formulate our analysis methods, involving the operator product expansion 
and the renormalization group in the framework of perturbative QCD.
In sect.~\ref{sec:results} we present our results, which include both BB and 3B 
potentials at both $\Nfv=2$ and 3.
A conclusion of this paper is given in sect.~\ref{sec:conclusion}.

\section{Basic idea and methods}
\label{sec:basics}
Since potentials themselves are scheme-dependent quantities 
like running couplings of QCD, we first give our definition (or scheme) for them. 
Let us consider the  equal-time NBS wave function~\cite{Balog:2001wv},
which in the case of a BB system for example is defined by
\begin{eqnarray}\label{wf2}
\varphi_E(\vec r) &=& \langle 0\vert B(\vec r, 0)B(0)\vert {\rm BB}, E\rangle_{\rm in}\,, 
\end{eqnarray}
where $\vert {\rm BB}, E\rangle_{\rm in}$ is a QCD asymptotic in-state with energy $E$ 
and $B(x)$ is a baryon interpolating operator made of 3 quarks such as 
$B(x) =\varepsilon^{abc}q^a(x)q^b(x)q^c(x)$ (flavor and spin labels suppressed).
Although a choice of $B(x)$ is rather arbitrary as long as it is coupled 
to a one baryon state, 
we take our choice to determine a scheme of the NBS wave function 
(like a choice of the renormalization scheme). 
It is important to note that the NBS wave function contains 
the information about the $BB$ scattering phase shift in QCD at 
energy $E$~\cite{Aoki:2005uf,Ishizuka:2009bx}.
Therefore, from this NBS wave function, the non-local but energy-independent BB 
potential is naturally defined so as to satisfy
\begin{eqnarray}
(e - H_0) \varphi_E(\vec r) &=& \int {\rm d}^3 r^\prime\, 
U(\vec r, \vec r^\prime) \varphi_E(\vec r^\prime)\,,
\quad e=\frac{k^2}{m_B},\ H_0=-\frac{\nabla^2}{m_B}\,,
\label{eq:SEQ}
\end{eqnarray} 
for all $E < E_{\rm th}$, where $E=2\sqrt{k^2+m_B^2}$ 
with a baryon mass $m_B$ and the threshold energy $E_{\rm th} = 2 m_B + m_M$ 
with a meson mass $m_M$. 
It is easy to show that the non-local but energy-independent 
$U(\vec r, \vec r^\prime)$ exists~\cite{Aoki:2009ji}. Indeed
\begin{eqnarray}
U(\vec r, \vec r^\prime) &=& \sum_{E,E^\prime <E_{\rm th}} (e-H_0) \varphi_E(\vec r)(\eta^{-1})^{EE^\prime}\varphi^*_{E^\prime}(\vec r^\prime)
\end{eqnarray} 
satisfies eq.~(\ref{eq:SEQ}) for all $E < E_{\rm th}$, 
where $(\eta^{-1})^{EE^\prime}$ is an inverse of
\begin{eqnarray}
\eta_{EE^\prime} &=& \int {\rm d}^3r\, \varphi_E^*(\vec r)\, \varphi_{E^\prime}(\vec r) 
\equiv (\varphi_E,\varphi_{E^\prime})
\end{eqnarray} 
in the restricted space where $E,E^\prime < E_{\rm th}$. 
One can construct $(\eta^{-1})^{EE^\prime}$ from eigenvectors corresponding 
to nonzero eigenvalues of the hermitian operator $\eta_{EE^\prime}$. 
Note that the non-local but energy-independent potential which satisfies eq.~(\ref{eq:SEQ})  
for all $E < E_{\rm th}$ is not unique, since one can add terms
involving wave functions for $E \ge E_{\rm th}$ which are orthogonal
to all wave functions with $E < E_{\rm th}$.

In practice, the non-local potential $U(\vec r,\vec r^\prime)$
is expanded in terms of the velocity as 
$U(\vec r,\vec r^\prime)= V(\vec r, \vec\nabla)\delta(\vec r -\vec r^\prime)$, 
whose lowest few orders for the NN case are given by
\begin{eqnarray}
V(\vec r,\vec\nabla) &=&V_0(r)+V_\sigma(r)\vec\sigma_1\cdot\vec\sigma_2 
+ V_T(r) S_{12}+V_{\rm LS}(r)\vec L\cdot\vec S +{\rm O}(\nabla^2)\,,
\end{eqnarray}
where $r=\vert \vec r\vert $, $\vec\sigma_i$ is the Pauli-matrix 
acting on the spin index of the $i$-th nucleon, 
$\vec S =(\vec\sigma_1+\vec\sigma_2)/2$ is the total spin, 
$\vec L=\vec r\times \vec p$ is the angular momentum, 
and $S_{12}$ is the tensor operator 
$S_{12}= 3(\vec r\cdot\vec\sigma_1)(\vec r\cdot\vec\sigma_1)/r^2-\vec\sigma_1\cdot\vec\sigma_2$.
This definition of the potential has been employed successfully 
to calculate NN potentials in 
lattice QCD for the first time~\cite{Ishii:2006ec,Aoki:2008hh,Aoki:2009ji},
where the central potentials at the leading order have 
qualitatively reproduced common features of phenomenological NN potentials: 
the force is attractive at medium to long distance 
while it has a short distance repulsion, the repulsive core.
After this success, the method has been widely applied to various hadronic 
interactions~\cite{Ishii:2009zr,Murano:2011nz,Nemura:2008sp,
Nemura:2012fm,Sasaki:2010bi,Ikeda:2011qm} 
including BB potentials at $\Nfv=3$~\cite{Inoue:2010hs,Inoue:2010es} 
and 3N forces at $\Nfv=2$~\cite{Doi:2011gq}. 
See ref.~\cite{aoki_review, Aoki:2012tk} for reviews of recent activities.  

In ref.~\cite{Aoki:2008yw} we made some toy model studies 
in order to better 
understand questions related to the scheme dependence, energy dependence, etc. of
the NBS wave functions and potentials. We noticed that in these 2-dimensional 
integrable models the short distance behavior of the NBS wave function
is represented very well by the 
leading terms appearing in the operator product expansion (OPE).
The hope is that the same qualitative feature applies to QCD,
for which, thanks to the property of asymptotic freedom,
the form of leading short distance behavior of the coefficient functions 
appearing in the OPE can be computed using perturbation theory also for
nucleons (and baryons). In this section we give a very short summary of 
our method using the example of the NN wave function and potential.
Our results for the cases NN~\cite{Aoki:2010kx}, BB~\cite{Aoki:2010uz},
3N~\cite{Aoki:2011aa} and 3B will be presented in the next section.

The behavior of the wave functions $\varphi_E(\vec r)$ 
at short distances ($r=\vert \vec r\vert\to0$) 
is encoded in the operator product expansion (OPE) of 
the two baryon operators: 
\begin{equation}
B(\vec r/2,0)\,B(-\vec r/2,0)\approx\sum_k
D_k(\vec r)\,{\cal O}_k(\vec 0,0)\,,
\end{equation}
where $\{ {\cal O}_k \}$ is a set of local color singlet 6-quark operators 
with two-baryon quantum numbers. It is important to note that
asymptotically the $\vec r$-dependence 
and energy dependence of the wave function is factorized into
\begin{equation}
\varphi_E(\vec r)\approx\sum_k D_k(\vec r)
\langle0\vert{\cal O}_k(\vec 0,0)\vert {\rm BB},E\rangle_{\rm in}\,.
\end{equation}

Now a standard renormalization group (RG) analysis gives~\cite{Aoki:2010kx}
the leading short distance behavior of the OPE coefficient function as
\begin{equation}
D_k(\vec r)\approx \lambda(r)^{-\nu_k}\,d_k\,,
\label{Dkasy}
\end{equation}
where $\lambda(r)$ is the 2-loop running coupling defined by
\begin{equation}
\frac{1}{\lambda}+\kappa\ln\lambda=\ln\frac{r_*}{r}\,,\,\,\,\,r\ll r_*\,,
\end{equation}
where (in the case of three light dynamical flavors) $\kappa=\frac{32}{81}$
and $r_*$ is some typical non-perturbative QCD radius 
(presumably O$(0.5$fm) but a precise
specification is not needed for our present considerations).

In (\ref{Dkasy}) $\nu_k$ is related to the 1-loop coefficient of the 
anomalous dimension of the operator ${\cal O}_k$, $d_k$ is the tree-level 
contribution of $D_k(\vec 0)$. Clearly the operator with largest
RG power~$\nu_k$ dominates the wave function (\ref{wf2}) at short distances.
We will denote the leading (largest) power by $\nu_1$ and the subleading
one (second largest) by $\nu_2$.

If neither $\nu_1$ nor $d_1$ vanishes, this leads to the leading
asymptotics of the s-wave potential of the form
\begin{equation}
V(r)\approx-\frac{\nu_1}{m_N\, r^2\left(\ln\frac{r_*}{r}\right)}\,,
\label{pot}
\end{equation}
which is attractive for $\nu_1>0$ and repulsive for $\nu_1<0$.
Note that although the running coupling itself is scheme dependent,
the above asymptotic form is not (and it is also energy independent).
Scheme dependence only affects the sub-leading contributions.
Whether (\ref{pot}) corresponds to an attractive or repulsive potential
is easily seen from the formula but can also be understood
intuitively from the short distance behavior of the corresponding
wave function. Namely, if $\nu_1>0$, the wave function diverges in the
origin (attraction), whereas for $\nu_1<0$ the wave function vanishes
at the origin (repulsion).

If $\nu_1=0$, the situation is more complicated. 
In this
case the relative sign of the ratio 
$R=\langle 0 \vert {\cal O}_2(\vec 0,0)\vert  {\rm NN}, E\rangle_{\rm in}/ \langle 0 \vert {\cal O}_1(\vec 0,0)\vert  {\rm NN}, E\rangle_{\rm in}$ 
between the leading contribution and the subleading 
contribution corresponding to an operator ${\cal O}_2$ with $\nu_2<0$
is important. As we will see this occurs in the NN case.
If $R$ is positive, the potential is repulsive, while it is
attractive for negative $R$. The above intuitive argument seems to work
also here, namely, for $R$ positive the wave function decreases approaching
the origin (repulsion) and the other way round for negative $R$.

To determine the asymptotic behavior we thus have to compute the 
spectrum of the renormalization group power $\nu_k$ corresponding to  
the operators appearing in the BB OPE.
The general form of a gauge invariant local 3--quark operator is given by
\beqa
B^F_\Gamma (x) \equiv B^{fgh}_{\alpha\beta\gamma}(x) 
= \varepsilon^{abc} q^{a,f}_\alpha(x) q^{b,g}_\beta(x) q^{c,h}_\gamma (x)\,,
\label{baryonop}
\eeqa
where $\alpha,\beta,\gamma$ are spinor, $f,g,h$ are flavor, 
$a,b,c$ are color indices of the (renormalized) quark field $q$.  
The color index runs from 1 to $N_c=3$, the spinor index from 1 to 4, 
and the flavor index from 1 to $\Nfv$. 
As indicated in (\ref{baryonop}) we use capital labels 
for sets of flavors $F=fgh$ and Dirac labels $\Gamma=\alpha\beta\gamma$.
Note that $B^{fgh}_{\alpha\beta\gamma}$ is symmetric under any
interchange of pairs of indices 
(e.g. $B^{fgh}_{\alpha\beta\gamma}=B^{gfh}_{\beta\alpha\gamma}$)
because the quark fields anticommute.
The usual nucleon operator which is employed in lattice simulations
is a certain linear combination constructed 
from the above operators~\cite{Aoki:2010kx}.

Gauge invariant local 6-quark operators can be written as 
linear combinations of simple operators 
$[BB]^{F_1F_2}_{\Gamma_1\Gamma_2}=B^{F_1}_{\Gamma_1}\,B^{F_2}_{\Gamma_2}$.
These local operators mix among themselves under renormalization. The
one-loop mixing matrix is given by diagrams corresponding to the exchange
of one gluon between all pairs of quark lines. Due to the flavor symmetry
(and partially to the chiral symmetry) of massless QCD, only those 6-quark
operators (of the form $[BB]^{F_AF_B}_{\Gamma_A\Gamma_B}$)
which preserve the set of flavor and Dirac indices as
\beqa
F_1\cup F_2&=&F_A\cup F_B, \quad
\Gamma_1\cup \Gamma_2=\Gamma_A\cup \Gamma_B\,,
\eeqa
occur in the mixing problem.
Note however that such operators are not all linearly independent.
We have the following \lq\lq gauge'' constraints:\footnote{We
call these constraints \lq\lq gauge'' constraints because we used them
to show the gauge invariance of the 1-loop anomalous dimensions.}
\beqa
3 [BB]^{F_1,F_2}_{\Gamma_1,\Gamma_2} + 
\sum_{i,j=1}^3 [BB]^{(F_1,F_2)[i,j]}_{(\Gamma_1,\Gamma_2)[i,j]} = 0\,.
\label{gaugeid2}
\eeqa
Here $[i,j]$ means a simultaneous exchange between the $i$-th indices of $F_1, 
\Gamma_1$ and the $j$-th indices of $F_2, \Gamma_2$.

The problem is thus reduced to finding the set of independent operators
after imposing the gauge constraints and then writing down and diagonalizing
the corresponding one-loop mixing matrix. This task can easily be automated
and we calculated and checked the results using independent Mathematica and Maple
programs. The size of the linear algebra problem turned out quite large
in some cases. For example the largest dimensions occur for a certain
set of Dirac indices in the BBB case: there are originally 14130 operators
before and 1518 after imposing the constraints. We are indebted to Thomas
Hahn for providing us with effective linear algebra tools for solving this 
part of the problem with Mathematica.

\section{Results}
\label{sec:results}

In the first three subsections we briefly summarize our results for 
the leading behavior of 2 nucleon, 2 baryon (involving operators with 
3 different flavors) and 3 nucleons; for the details of these 
computations the reader is requested to consult 
refs.~\cite{Aoki:2010kx,Aoki:2010uz,Aoki:2011aa} respectively.
Our latest results concerning the case of 3 baryon operators 
are presented in the fourth subsection.

In each case the computation consists of three steps.
i) a basis of independent gauge invariant operators of engineering dimension
equal to that of the operator product appearing in the definition of the BS 
wave function is determined for each Dirac structure. 
ii) The spectrum of 1-loop anomalous dimensions $\gamma_k$ of the linear combination 
of these operators which do not mix under renormalization (at 1-loop) are computed.
iii) The operators appearing in the OPE at tree level are listed.

For operators which renormalize multiplicatively (without mixing)
\beqa
O^{\mathrm{ren}}=Z_O(\epsilon,g) O^{\mathrm{bare}}\,,
\eeqa
the associated anomalous dimensions are, using dimensional regularization in 
$D=4-2\epsilon$ dimensions, specified by 
\beqa
\gamma_O = -\left[\epsilon g + \beta_0 g^3 +\dots\right]\frac{\partial Z_O}{\partial g}
= \gamma_{O0} g^2 +\mathrm{O}(g^4)\,,
\eeqa
where $\beta_0$ is the QCD 1-loop beta function coefficient
\beqa
\beta_0=\frac{1}{16\pi^2}\left[11-\frac23\Nf\right]\,.
\eeqa
We will restrict attention to the octet baryon operators appearing 
in the definition of the BS wave functions, which have 1-loop anomalous 
dimension 
\beqa  
\gamma_{B0}= 24 d\,,\,\,\,\,\,\,\,\,d\equiv 1/(96\pi^2) \,.   
\eeqa
It turns out that all 1-loop anomalous dimensions of operators 
appearing below can be expressed as even integers times $2d$.

\subsection{The NN potentials}

The computationally easiest case is the BS wave function defined 
with the product of two nucleons, operators defined with just two ($u,d$) 
flavors of quarks.
The spectrum of anomalous dimensions in this case has a striking structure:
\beqa  
\gamma_{k0}\le 2\gamma_{B0}\,,\,\,\,\,\,\forall k\,.   
\eeqa
It is easy to see that there are NN (also BB) operators with $\gamma_{k0}= 2\gamma_{B0}$
- e.g. those of the form $B_+B_-$ where $B_\pm$ involves only 
quarks of positive (negative) chirality. 
Such operators generically appear in the OPE at tree level 
and hence in the NN case these dominate the short distance 
behavior of the BS wave function. 
However since $D_k=const.$ asymptotically for these operators, 
their contributions alone 
do not determine the short distance behavior of the potentials 
(defined through taking derivatives of the wave function). 
The leading behavior of the (central) potentials involves also
the next-to-largest anomalous dimensions, which are found to be
$\gamma^{(0,1)}=24 d\,,\gamma^{(1,0)}=40 d$
in the two spin-isospin channels $(S,I)=(0,1),(1,0)$ respectively:
\beqa  
V_c^{(S,I)}(r)\sim R^{(S,I)} r^{-2}(-\ln r)^{\beta^{(S,I)}-1}\,,  
\label{VtwoN}
\eeqa
with (assuming $\Nf$ is such that $\beta_0>0$)
\beqa
\beta^{(S,I)}=\frac{\gamma^{(S,I)}-2\gamma_{B0}}{2\beta_0}< 0\,,
\eeqa
and the constants $R^{(S,I)}$ appearing above involve
the ratio of matrix elements of the leading and subleading operators in the OPE as
\beqa
R^{(S,I)} =\frac{\langle 0 \vert {\cal O}_2^{(S,I)}\vert  {\rm NN}, E\rangle_{\rm in}}{\langle 0 \vert {\cal O}_1^{(S,I)}\vert  {\rm NN}, E\rangle_{\rm in}} .
\eeqa
Thus although the OPE determines the functional form of the short distance 
behavior, unfortunately it does not determine the sign  in this special case. 
Information on the sign of $R^{(S,I)}$ requires additional non-perturbative input,
which could e.g. be supplied by dedicated numerical simulations. 
In~\cite{Aoki:2010kx} we have given arguments based on a non-relativistic 
expansion that the $R^{(S,I)}$ are positive, 
in which case the potential is repulsive in both channels
as indeed observed in the numerical simulations.

\subsection{The BB potentials}

We next consider the operator product of two baryon octet operators
where the operators appearing in the OPE can involve 3 different flavors.
The tensor product of two octets can be decomposed into 6 channels, according to
\beqa  
8\otimes 8= (1\oplus 8\oplus 27)_s+(8\oplus 10 \oplus\overline{10})_a\,,
\eeqa
where subscripts $s$($a$) represents the property that the operator is (anti-)symmetric under the interchange of two baryons.
The two nucleon states with $(S,I)=(0,1)$ or $(1,0)$  considered above
belong to the $27_s$ or  $\overline{10}_a$ representations, respectively.
It is sufficient to consider the situation of composite 6-quark
operators with two quarks of each flavor ($uuddss$) since all 6 SU(3) 
representations appear in this type of operator.

Although most of the operators have 1-loop anomalous dimensions $\le 2\gamma_{B0}$
we found also operators with anomalous dimensions greater than $2\gamma_{B0}$ 
corresponding to attractive BS potentials in the three channels 
$1_s,8_a,8_s$; they are listed in Table~\ref{tab:gamma2B}. 
\begin{table}[tb]
\caption{List of channels with anomalous dimensions greater than $2\gamma_{B0}$;
others can be obtained by symmetry transforms of the Dirac indices $1\leftrightarrow 2\,,
3\leftrightarrow 4$ or $(1,2)\leftrightarrow(3,4)$.}
\label{tab:gamma2B}
\begin{center}
\begin{tabular}{|c|c|c|}
\hline
Dirac structure & $\gamma^{\mathcal{R}}_0/(2d)$ & SU(3) representation \\
\hline
$112334$ & $42$ & $1$\\
$111222$ & $36$ & $1$\\
$112234$ & $36$ & $1$\\
$111234$ & $32$ & $1\oplus 8$\\
$112234$ & $32$ & $1\oplus 8$\\
\hline
\end{tabular}
\end{center}
\end{table}
There are no attractive operators in the $27_s,10_a,\overline{10}_a$ channels,
which is consistent with what we found in the two nucleon case.
All the attractive operators appear in the OPE of two
baryon operators at tree level and therefore 
the corresponding BS potentials in these channels (denoted by 
$\mathcal{R}$)  
behave asymptotically as in eq.~(\ref{pot}) with
\beqa
\nu_1=\frac{\gamma^{\mathcal{R}}_0-2\gamma_{B0}}{2\beta_0}>0\,.
\label{2bres}
\eeqa
The larger anomalous dimensions occur in the lower dimensional SU(3)
representations. Indeed the tendency of repulsion among the operators 
with fixed total number of valence quarks is increased if the number
of participating valence flavors is decreased; this is a first indication
of what we suggestively call a ``Pauli suppression principle" at work.

Monte Carlo simulations in 3-flavor lattice QCD~\cite{Inoue:2010hs, Inoue:2010es}
have as yet only found attraction in the $1_s$ channel. 
The fact that attraction in octet channels has not yet been 
observed may have many different reasons. Firstly we note that the
amplitudes of attractive operators in the $8_s$ channel vanish 
in the non-relativistic limit.
Secondly, and this comment applies to all our considerations,
we do not know at which distance $r$ the asymptotic formulae (\ref{pot})
set in. The physical distances of the BS potentials probed 
in Monte Carlo simulations up to now ($>\sim 0.2$ fm) may not be small enough. 

\subsection{The 3N potentials}

For the investigation of the 3 baryon potentials we consider 
equal-time NBS wave functions involving the product of three baryon operators:
\begin{eqnarray}\label{wf3}
\psi_{{\rm 3B},E}(\vec r, \vec \rho) &=& 
\langle 0 \vert B({\vec x}_1, 0) B({\vec x}_2, 0)B({\vec x}_3, 0) \vert  {\rm 3B}, E\rangle_{\rm in}\,,
\end{eqnarray}
where $\vert  {\rm 3B}, E\rangle_{\rm in}$ is a 3-baryon asymptotic in-state of energy $E$. 
In the argument of $\psi_{{\rm 3B},E}$ we have introduced Jacobi coordinates 
$\vec r={\vec x}_1-{\vec x}_2\,, 
\vec \rho=\left[{\vec x}_3-({\vec x}_1+{\vec x}_2)/2\right]/\sqrt{3}$.
The three baryon potential is defined through the wave functions by
\beqa
&&\left[\frac{1}{m_B}\left(\nabla^2_r+\nabla^2_\rho\right)+\mathcal{V}_{\rm 3B}\right]
\psi_{{\rm 3B},E}(\vec r, \vec \rho)=E\psi_{{\rm 3B},E}(\vec r, \vec \rho)\,,
\\
&&\mathcal{V}_{\rm 3B}(\vec r, \vec \rho)
=\sum_{1\le i<j\le3}V_{\rm BB}({\vec x}_i-{\vec x}_j)+V_{\rm 3B}(\vec r, \vec \rho)\,,
\label{VthreeB}
\eeqa
where the reduced masses of these Jacobi coordinates are $m_B/2$
and only the leading contributions in the velocity expansion are presented. 
The $V_{\rm BB}$ are the BB potentials defined through the BB wave functions.

The leading terms in the OPE of 3-baryon operators in $\vec r, \vec \rho\to0$ limits involves the associated local
gauge invariant 9-quark operators (of minimal engineering dimension of 27/2).

For the 3N case the spectrum of anomalous dimensions 
of the the 9 quark operators (now involving only two flavors) 
have the simple property:
\beqa  
\gamma_{A0}< 3\gamma_{B0}\,,\,\,\,\,\,\forall A\,.   
\eeqa
Again it turns out that most of the operators with largest anomalous dimensions 
appear in the OPE at tree level. From this result, without recourse to additional nonperturbative arguments, 
we deduce in this case that generically the total central 3N potential $\mathcal{V}_{\rm 3N}(\vec r, \vec \rho)$
is repulsive at short distances:
\beqa  
\mathcal{V}_{\rm 3N}(\vec r, \vec \rho)\simeq \frac{-4\beta_A}{m_N s^2(-\ln(s/r_*))}  
\label{calVthreeN}
\eeqa
where $s=\sqrt{(\vec r)^2+(\vec \rho)^2}$ and
\beqa
\beta_A=\frac{\gamma_{A0}-3\gamma_{B0}}{2\beta_0}<0\,,\,\,\,\,\,\max_{A}\gamma_{A0}=16d\,.
\label{3bres}
\eeqa
Since the singularity of $V_{\rm NN}$ given in (\ref{VtwoN}) is milder than 
that of $\mathcal{V}_{\rm 3N}$ in (\ref{calVthreeN}), it follows that the
3N potential $V_{\rm 3N}$ defined in (\ref{VthreeB}) (with B $\to$ N) is responsible
for the asymptotic repulsion.

We note that increasing the number of quarks, but not the number of flavors,
tends to decrease the anomalous dimensions of the operators, 
thereby increasing the tendency for a repulsive core of the potential,
which is a further indication of a ``Pauli suppression principle" mentioned before.

\subsection{The 3B potentials}

This problem is very similar to the previous three cases but since the
corresponding results have not yet been published we now describe
the outcome of our considerations in some detail. However, due to the
very large dimension of the problem we refrain from giving 
full details in the paper.
The tensor product of three octets can be decomposed as
\beqa
8\otimes 8\otimes  8 &=&
 64 \oplus ( 35 \oplus \overline{35})_2 \oplus  27_6 \oplus ( 10\oplus \overline{10})_4 \oplus 8_8 \oplus 1_2 
\eeqa
where the subscript indicates the multiplicity of the corresponding irreducible representation.

In the three-baryon case we are considering 9-quark operators with
three quarks of each flavor ($uuudddsss$). 
We find that there are 35 different Dirac 
structures (plus their transforms under the symmetries $1\leftrightarrow2$,
$3\leftrightarrow4$ or $(1,2)\leftrightarrow(3,4)$). The simplest Dirac 
structure is $111111111$ and there are 9 such operators 
but obviously only one of them (corresponding to 
$\prod_{a=1}^3(u^a_1d^a_1s^a_1)$) 
remains after imposing the gauge constraints. The anomalous dimension of this
operator is $-36$ (in $2d$ units) and is an SU$(3)$ singlet. The next simplest
case corresponds to the Dirac structure $111111112$ where there are 
originally 51 operators but only 3 remain independent after imposing the 
gauge constraints. The remaining three 
operators correspond to anomalous dimension $-36$ and form a singlet and
the $I=0$ and $I=1$ members of an SU$(3)$ octet. Next we consider the Dirac
case $111111122$. Here the number of operators is reduced from 153 to 9.
The results are summarized in Table~\ref{tab:table2}. 
\begin{table}[tb]
\caption{The Dirac case $111111122$.}
\label{tab:table2}
\begin{center}
\begin{tabular}{|c|c|}
\hline
$\gamma_{A0}/(2d)$ & SU(3) representation \\
\hline
$-52$ & $27$\\
$-36$ & $1,8$\\
$-22$ & $8$\\
$-4$ & $1$\\
\hline
\end{tabular}
\end{center}
\end{table}
For more complicated Dirac
structures the number of operators increases rapidly and for example for
$111333224$ and $111222334$ we have 9960 operators, which are reduced to 1014
by the gauge constraints. The largest case is $111223344$. The number of 
operators here is reduced from 14130 to 1518. The spectrum of these 1518 independent
operators is
\begin{center}$
\{-76,-60,-54,-52,-48,-46,-44,-42,-40,-36,-34,$

$-30,-28,-24,-22,-20,-18, -16,-12,-10,-8,-6,-4,$

$0,2,4,6,8,12,14,18,20,24,26,30,32,36,38,42,44,48\}$
\end{center}
and the dimensions of the SU$(3)$ representations in which these operators
transform are
\begin{center}$
\{1,8,10,27,28,35,55,64,80,81\}$\,.
\end{center}
All eigenvalues and the corresponding SU$(3)$ representations are
given in Table~\ref{tab:monster} for this case. 

The most important new feature of the spectrum of operators with
three-baryon quantum numbers (with respect to the three-nucleon case) is the
presence of attractive channels, though the vast majority of eigenvalues
correspond to repulsion as before. We give a complete list of such
channels (corresponding to eigenvalues $\gamma_{A0}\ge 3\gamma_{B0}$) in 
Table~\ref{tab:table3}, where an asterisk superscript ($^*$)
indicates if the operator is already present at tree level in the product 
of three baryon operators.

As can be seen from the table, we have found tree operators with 
$\gamma_{A0}> 3\gamma_{B0}$ in the $1$ (singlet) and $8$ (octet) channels,
which give a universal attraction at short distances (attractive cores).   
In addition, there are tree level operators with 
$\gamma_{A0} = 3\gamma_{B0}$ in the $8$ (octet) and $10, \overline{10}$ 
(decuplet) channels, which lead to non-universal attraction or repulsion at 
short distances depending on the sign of the ratio of associated matrix 
elements. On the other hand, universal repulsion at short distances 
(repulsive core) appears only in the $64, 35, \overline{35}, 27$ channels.
\begin{table}[tb]
\caption{List of channels with anomalous dimensions greater than or equal to 
$3\gamma_{B0}$. Operators present at tree level in the OPE are indicated
by an asterisk $^*$.}
\label{tab:table3}
\begin{center}
\begin{tabular}{|c|c|c|}
\hline
Dirac structure & $\gamma_{A0}/(2d)$ & SU(3) representation \\
\hline
$111223344$ & $48$ & $1,8^*$\\
            & $44$ & $1^*,8^*$\\
            & $42$ & $1^*,8^*$\\
            & $38$ & $1,8$\\
            & $36$ & $8^*, 10^*, \overline{10}^*$ \\
$111333224$ & $44$ & $1^*,8^*$\\
            & $38$ & $1,8$\\
            & $36$ & $1$ \\
$111222334$ & $48$ & $8^*$\\
            & $44$ & $1^*$\\
$111133442$ & $42$ & $1^*,8^*$\\
            & $38$ & $1,8$\\
$111122334$ & $44$ & $1^*$\\
                           & $36$ & $1$ \\
$111133324$ & $38$ & $1,8$\\
$111113324$ & $36$ & $1$ \\
\hline
\end{tabular}
\end{center}
\end{table}

Note however, that the anomalous dimensions listed in the table give,
using (\ref{calVthreeN}) and (\ref{3bres}), the asymptotic behavior 
of the \lq\lq total'' 3B potential $\mathcal{V}_{\rm 3B}$ and not the 
\lq\lq true'' 3B potential $V_{\rm 3B}$, which is obtained from the 
former by subtracting the sum of the 2B potentials. In the 3N case 
this distinction was not important, since the leading behavior of the 
2N potentials is always milder than that of the 3N potential,
which is hence the dominant contribution at short distances. 
The 3B case is made more complicated by the fact that both the
\lq\lq total'' 3B potential $\mathcal{V}_{\rm 3B}$ and the
2B potential $V_{\rm 2B}$ have some attractive channels, competing
with each other.
  
Take for example the 3B configuration with Dirac structure 111223344,
which contains the largest anomalous dimension 48 (in $2d$ units).
This is in the octet channel and leads to an attractive total force, 
whose strength is proportional
to the numerator of the formula (\ref{3bres}), 12 in our units.
On the other hand, there is a 2B contribution to this total force
coming from two baryons with Dirac structure 112334 and forming a
singlet state with the highest anomalous dimension 42 in our units (times a free octet baryon).
From (\ref{2bres}) we see that the corresponding 2B force is proportional
to 18. So in this example the 2-body attractive potential is actually 
stronger than the total 3-body attraction hence there must be a repulsive 
effect coming from $V_{\rm 3B}$ as
\beqa
V_{\rm 3B}(\vec r,\vec\rho) &\simeq& \frac{1}{m_B}
\left[\frac{-4\beta_{\rm 3B}^{\rm max}}{s^2(-\ln(s/r_*))}
+\sum_{1\ge i \ge j\ge 3}\frac{\beta_{\rm BB}^{\rm max}}
{r_{ij}^2(-\ln(r_{ij}/r_*))}\right] 
\eeqa
as $\vert \vec r\vert, \vert \vec \rho \vert\rightarrow 0$,
where $m_B$ is the octet baryon mass\footnote{We here ignore mass differences among octet baryons caused by the flavor SU(3) breaking.}, and ( for $N_f=3$)
\beqa
\beta_{\rm 3B}^{\rm max} &=& \frac{2}{9},\quad  \beta_{\rm BB}^{\rm max} = \frac{1}{3},\quad
s=\sqrt{\vec r^2+\vec \rho^2}, \\
r_{12}&=&\vert \vec r\vert, \quad r_{13}=\vert \vec r/2 -\sqrt{3}\vec\rho\vert, \quad r_{23}=\vert \vec r/2 +\sqrt{3}\vec\rho\vert .
\eeqa
As an other similar example we take the Dirac structure 111133442
in the singlet channel with anomalous dimension 42. A two-baryon operator
with Dirac structure 111234 in the octet channel
(combined with an extra octet baryon)
contributes to this component of the 3B force and the corresponding numbers are:
\beqa
\beta_{\rm 3B}^{\rm max} &=& \frac{1}{9},\qquad  \beta_{\rm BB}^{\rm max} = \frac{4}{27}.
\eeqa

We do not further attempt to analyze the general case, which is clearly very
complex since in addition to the cancellation between a 2B and a 3B force
as in the above examples, there are cases where cancellation occurs among
the 2B forces.

\section{Conclusion}
\label{sec:conclusion}

In this paper, we have presented our recent activities on determinations 
of short distance behaviors of BB and 3B potentials
defined from the NBS wave function in QCD 
using the operator product expansion and the RG analysis in perturbative QCD.
Our results show that the repulsion at short distance (repulsive core) 
become stronger for more participating valence quarks 
or less different numbers of flavors $\Nfv$. 
Explicitly we have: 
\begin{enumerate}
\item NN potentials ($\Nfv=2$) seem to have a repulsive core at low energy, 
which however is not universal in the sense that the coefficient may 
depend on the properties of the NBS wave function such as total energy.
\item  BB potentials ($\Nfv=3$) can have not only a repulsive core but also 
channels with an attractive core, the latter of which is universal 
({\it i.e.} energy independent).
In particular, the attractive core in the flavor singlet potential 
has indeed been confirmed in lattice QCD simulations~\cite{Inoue:2010hs,Inoue:2010es}. 
\item 3N potentials have a universal repulsive core. 
This is the most unambiguous result in our project.
\item In the 3B case we were able to study the short distance asymptotics
of the \lq\lq total'' 3B potential $\mathcal{V}_{\rm 3B}$ only with our methods.
Although the vast majority of various channels ($64, 35, \overline{35}, 27$) 
in  3B potentials have a universal repulsive core, 
a few channels ($1,8$) have a universal attractive core. 
In addition, there may appear a non-universal repulsive or 
attractive core in the $10,\overline{10}$ channels. 
\end{enumerate}
Roughly speaking, these results suggest that a 
``Pauli suppression principle" among quarks is at work.
Explicit 1-loop calculations are needed, however, 
to obtain the detailed structure of the 
anomalous dimensions of BB or 3B operators.

Our results give important information on the short distance 
behaviors of baryonic interactions, 
which can be used to constrain baryonic potentials at 
short distance obtained in lattice QCD~\cite{Aoki:2012tk,Aoki:2008yw}, 
in particular, to constrain 3B potentials~\cite{Doi:2011gq} .
As mentioned in the introduction, the universal repulsion of 3B 
potentials seems necessary to explain an existence 
of two-solar-mass neutron star. 
Our result shows that the vast majority of 3B channels have a repulsive core 
in the 3B potential, though a few channels have an attractive core. 
It is interesting and important to check these qualitative behaviors 
of 3B potentials at short distances 
and furthermore to investigate the strength of these cores, 
employing lattice QCD simulations.

\section*{Acknowledgments}

S. A would like to thank Dr. T. Doi  for 
useful discussions. We thank T. Hahn for helping with Mathematica
linear algebra.
S. A. is supported in part by Grant-in-Aid for Scientific Research on 
Innovative Areas (No. 2004: 20105001,20105003) and by 
SPIRE (Strategic Program for Innovative Research).
This investigation was also supported in part by the Hungarian National 
Science Fund OTKA (under K83267). 
S. A. and J. B. would like to thank the Max-Planck-Institut f\"ur
Physik for its kind hospitality during their stay for this research project.

\begin{table}[bh]
\caption{The Dirac case $111223344$. Operators present at tree level 
in the OPE are indicated by an asterisk $^*$.}
\label{tab:monster}
\begin{center}
\begin{tabular}{|c|c|}
\hline
$\gamma^{\mathcal{R}}_0/(2d)$ & SU(3) representation (multiplicity)\\
\hline
$48$ & $1,8^*$ \\
$44$ & $1^*,8^*$ \\
$42$ & $1^*,8^*$ \\
$38$ & $1,8$ \\
$36$ & $8^*,10^*$ \\
$32$ & $1^*,8^*(3),10^*(2)$ \\
$30$ & $8,10^*$ \\
$26$ & $1,8(3),10(2)$ \\
$24$ & $1,8^*(4),10^*(3),27^*(2)$ \\
$20$ & $1^*(2),8^*(6),10^*(4),27^*(3)$ \\
$18$ & $1,8^*(4),10^*(3),27^*(2)$ \\
$14$ & $1(2),8(5),10(3),27(3)$ \\
$12$ & $1(3),8^*(7),10^*(5),27^*(4),35^*$ \\
$8$ & $1,8^*(8),10^*(9),27^*(5),35^*(2)$ \\
$6$ & $1^*(2),8^*(6),10^*(6),27^*(4),35^*(2)$ \\
$4$ & $1,8(3),10(2),27$ \\
$2$ & $8(6),10(9),27(5),35(3)$ \\
$0$ & $1(3),8^*(9),10^*(10),27^*(9),35^*(5),64$ \\
$-4$ & $1^*(4),8^*(9),10^*(10),27^*(9),35^*(6),64$ \\
$-6$ & $1(2),8^*(8),10^*(9),27^*(7),35(4),64$ \\
$-8$ & $8(2),10(3),27,35$ \\
$-10$ & $1(2),8(7),10(9),27(8),35(6),64(2)$ \\
$-12$ & $1,8(5),10(8),27(5),35(5),64$ \\
$-16$ & $1(3),8^*(9),10^*(11),27^*(11),28,35^*(9),64(3)$ \\
$-18$ & $1,8^*(7),10^*(12),27^*(9),28,35^*(10),64(3),81$ \\
$-20$ & $1,8(4),10(5),27(4),35(3),64$ \\
$-22$ & $1,8(5),10(9),27(8),28(2),35(9),64(2),81$ \\
$-24$ & $1,8^*(4),10(5),27^*(7),28(4),35^*(9),64(3),81$ \\
$-28$ & $1(2),8(6),10(7),27(7),28(2),35(9),64(4),81(2)$ \\
$-30$ & $1,8(4),10(6),27(7),28(2),35(8),64(3),81$ \\
$-34$ & $1,8(5),10(6),27(6),28,35(7),64(4),81(2)$ \\
$-36$ & $8(3),10(4),27(4),28,35(5),64(3),81(2)$ \\
$-40$ & $8,10(3),27(3),28(2),35(5),64,81(2)$ \\
$-42$ & $8(2),10(3),27(3),28,35(4),64(2),81(2)$ \\
$-44$ & $8,10,27,35,64,81$ \\
$-46$ & $8,10(3),27(3),28(2),35(5),64(2),80,81(2)$ \\
$-48$ & $27,28,35,64,80,81$ \\
$-52$ & $10,27(2),28(2),35(3),64(2),80(2),81(2)$ \\
$-54$ & $10,27,28,35(2),64,80,81$ \\
$-60$ & $10,27(2),28(2),35(3),64(2),80(2),81(2)$ \\
$-76$ & $35,55,64,80,81$ \\
\hline
\end{tabular}
\end{center}
\end{table}



\begin{thebibliography}{99}
\bibitem{NN-review}
M.~Taketani et al., Prog. Theor. Phys. Suppl. {\bf 39}, 1-346  (1967),
N.~Hoshizaki et al., ibid. {\bf 42}, 1-159 (1968). \\ 
G.~E.~Brown and A.~D.~Jackson, {\em Nucleon-nucleon Interaction},
 (North-Holland, Amsterdam, 1976).\\
R.~Machleidt, Adv.\ Nucl.\ Phys.\  {\bf 19}, 189 (1989).\\
R.~Machleidt and I.~Slaus, {\it J. Phys.} {\bf G27}, R69 (2001).

\bibitem{yukawa}
H.~Yukawa, Proc. Math. Phys. Soc. Japan, {\bf 17}, 48 (1935).

\bibitem{jastrow}
R.~Jastrow, Phys. Rev. {\bf 81}, 165 (1951).

\bibitem{Demorest:2010bx}
P.~Demorest, T.~Pennucci, S.~Ransom, M.~Roberts and J.~Hessels,
Nature {\bf 467}, 1081 (2010)
[arXiv:1010.5788 [astro-ph.HE]].

\bibitem{Nishizaki:2002ih}
S.~Nishizaki, T.~Takatsuka and Y.~Yamamoto,
Prog.\ Theor.\ Phys.\  {\bf 108}, 703 (2002).\\
T.~Takatsuka, S.~Nishizaki and R.~Tamagaki,
Prog.\ Theor.\ Phys.\ Suppl.\  {\bf 174}, 80 (2008).

\bibitem{Ishii:2006ec}
N.~Ishii, S.~Aoki and T.~Hatsuda,
Phys.\ Rev.\ Lett.\  {\bf 99}, 022001 (2007) [arXiv:nucl-th/0611096].

\bibitem{Aoki:2008hh}
S.~Aoki, T.~Hatsuda and N.~Ishii,
Comput.\ Sci.\ Dis.\  {\bf 1}, 015009 (2008)
 [arXiv:0805.2462 [hep-ph]].

\bibitem{Aoki:2009ji}
S.~Aoki, T.~Hatsuda and N.~Ishii,
Prog.\ Theor.\ Phys.\ {\bf 123}, 89(2010)
 [arXiv:0909.5585 [hep-lat]].
  
\bibitem{Ishii:2009zr}
N.~Ishii, S.~Aoki and T.~Hatsuda,
PoS LATTICE2008, 155 ( 2008)
 [arXiv:0903.5497 [hep-lat]].

\bibitem{Murano:2011nz} 
K.~Murano, N.~Ishii, S.~Aoki and T.~Hatsuda,
 Prog.\ Theor.\ Phys.\  {\bf 125}, 1225 (2011) [arXiv:1103.0619 [hep-lat]].

\bibitem{Nemura:2008sp}
H.~Nemura, N.~Ishii, S.~Aoki and T.~Hatsuda,
Phys.\ Lett.\  B {\bf 673}, 136 (2009)
 [arXiv:0806.1094 [nucl-th]].

\bibitem{Nemura:2012fm} 
H.~Nemura and f.~H.~Q.~Collaboration,
arXiv:1203.3320 [hep-lat].

\bibitem{Sasaki:2010bi} 
  K.~Sasaki [for HAL QCD Collaboration],
  PoS LATTICE {\bf 2010}, 157 (2010)
  [arXiv:1012.5685 [hep-lat]].

\bibitem{Ikeda:2011qm} 
  Y.~Ikeda [for HAL QCD Collaboration],
  arXiv:1111.2663 [hep-lat].
   
\bibitem{Inoue:2010hs}
  T.~Inoue {\it et al.}  [HAL QCD collaboration],
Prog.\ Theor.\ Phys.\ {\bf 124}, 591 (2010)
  [arXiv:1007.3559 [hep-lat]].

\bibitem{Inoue:2010es}
  T.~Inoue {\it et al.}  [HAL QCD Collaboration],
  Phys.\ Rev.\ Lett.\  {\bf 106}, 162002(2011)
 [ arXiv:1012.5928 [hep-lat]].

\bibitem{Doi:2011gq} 
  T.~Doi, S.~Aoki, T.~Hatsuda, Y.~Ikeda, T.~Inoue, N.~Ishii, K.~Murano and H.~Nemura {\it et al.},
Prog.\ Theor.\ Phys.\ {\bf 127},  723 (2012)
  [arXiv:1106.2276 [hep-lat]].

\bibitem{Aoki:2010kx}
  S.~Aoki, J.~Balog and P.~Weisz,
JHEP05, 008 (2010) 
  [arXiv:1002.0977 [hep-lat]].

\bibitem{Aoki:2009pi}
  S.~Aoki, J.~Balog and P.~Weisz,
PoS LAT2009, 132 (2009)
  [arXiv:0910.4255 [hep-lat]].

\bibitem{Aoki:2011aa}
  S.~Aoki, J.~Balog and P.~Weisz,
  New J.\ Phys.\  {\bf 14} (2012) 043046
  [arXiv:1112.2053 [hep-lat]].

\bibitem{Aoki:2010uz}
  S.~Aoki, J.~Balog and P.~Weisz,
  JHEP09, 083 (2010)  [arXiv:1007.4117 [hep-lat]].

\bibitem{Balog:2001wv} 
  J.~Balog, M.~Niedermaier, F.~Niedermayer, A.~Patrascioiu, E.~Seiler and P.~Weisz,
  Nucl.\ Phys.\ B {\bf 618}, 315 (2001)
  [hep-lat/0106015].

\bibitem{Aoki:2005uf}
  S.~Aoki {\it et al.}  [CP-PACS Collaboration],
  Phys.\ Rev.\ D {\bf 71} (2005) 094504
  [hep-lat/0503025].

\bibitem{Ishizuka:2009bx}
  N.~Ishizuka,
  PoS LAT {\bf 2009} (2009) 119
  [arXiv:0910.2772 [hep-lat]].
  

\bibitem{aoki_review}
S.~Aoki,
Prog.\ Part.\ Nucl. Phys.\ {\bf 66}, 687  (2011) 
[arXiv:1107.1284 [hep-lat] ].

\bibitem{Aoki:2012tk}
  S.~Aoki {\it et al.}  [HAL QCD Collaboration],
  arXiv:1206.5088 [hep-lat].

\bibitem{Aoki:2008yw}
S.~Aoki, J.~Balog and P.~Weisz,
Prog.\ Theor.\ Phys.\  {\bf 121}, 1003 (2009)
[arXiv:0805.3098 [hep-th]].

\end{thebibliography}
\end{document}